\documentclass[12pt]{iopart}

\usepackage{pstricks,graphicx,bbm,amssymb}
\usepackage[utf8]{inputenc}
\usepackage[english]{babel}
\usepackage{color}
\usepackage{cite}

\newcommand{\erf}{\mathrm{erf} \, }

\def\inf{{\infty}}
\def\Id{{\mathbbm 1}}

\usepackage{xparse}
\NewDocumentCommand{\ceil}{s O{} m}{%
  \IfBooleanTF{#1} 
    {\left\lceil#3\right\rceil} 
    {#2\lceil#3#2\rceil} 
}

\begin{document}
\title[Hybrid near-optimum binary receiver with realistic$\ldots$]{Hybrid near-optimum binary receiver with realistic photon-number-resolving detectors}

\author{M.~N.~Notarnicola, M.~G.~A.~Paris and S.~Olivares}
\address{Dipartimento di Fisica ``Aldo Pontremoli'',
Universit\`a degli Studi di Milano and INFN Sezione di Milano, via Celoria 16, I-20133 Milano, Italy}
\ead{stefano.olivares@fisica.unimi.it}

\begin{abstract}
We propose a near-optimum receiver for the discrimination of binary phase-shift-keyed coherent states employing photon-number-resolving detectors. The receiver exploits a discrimination strategy based on both the so-called homodyne-like and the direct detection, thus resulting in a hybrid scheme. We analyse the performance and the robustness of the proposed scheme under realistic conditions, namely, in the presence of inefficient detection and dark counts. 
We show that the present hybrid setup is near-optimum and beats both the standard-quantum-limit and the performance of the Kennedy receiver.
\end{abstract}
\maketitle

\section{Introduction}
The problem of discriminating quantum states is a challenging task in quantum information theory, since quantum mechanics does not allow perfect discrimination if the considered states are not orthogonal. In particular, the task of coherent state discrimination is of great relevance for quantum communications, since these states are the typical information carrier in optical channels, finding a large application both in physics and telecommunication engineering \cite{Cariolaro}.
The simplest scenario is the binary phase shift keying (BPSK), where one has to discriminate between two coherent states with the same energy but a $\pi$ phase difference \cite{Cariolaro,Helstrom,Bergou}.
In this case the theory developed by Helstrom \cite{Helstrom, Bergou} identifies the minimum error probability, the so-called Helstrom bound, rising the question concerning the possible implementation of an optimal receiver able to achieve this bound.
\par
Several proposals of feasible optimum or near-optimum receivers have been advanced in literature, based on either single shot discrimination or feedback-based strategies. As regards single-shot strategies, there are several options. Homodyne receivers \cite{Cariolaro} are constructed as an extension of the classical systems for discrimination of signals and are based on the measurement of the quadratures of the optical field. The Kennedy receiver \cite{KennedyR} is based on a nulling displacement operation followed by photodetection and proves to be near-optimum, reaching in the high energy regime twice the Helstrom bound. Recently such a scheme has also been improved by Takeoka and Sasaki \cite{ImprovedKennedy} by optimizing the displacement operation , obtaining a further advantage in the range of small energies. Finally, Sasaki and Hirota \cite{Sasaki&Irota} have proposed a scheme which is able to reach the Helstrom bound based on the application of unitary operations defined in the two-dimensional space spanned by the coherent states. However, the realization of such unitary would require highly non-linear optical elements, making this kind of receiver not realizable with the usual practical linear optics components.
\par
Better results are obtained with feedback strategies. Dolinar \cite{Dolinar} extended the principle of the Kennedy receiver by designing a new receiver employing a time-varying displacement operation conditioned on the outcome of continuous photodetection. The Dolinar receiver is indeed optimum and reaches the Helstrom bound. More recently, the Dolinar approach has been revised and less demanding strategies have been proposed which employ feed-forward methods exploiting the \textit{slicing} of the coherent state. In these discrimination strategies, the incoming state is split into a finite number of copies with smaller energies and each copy is measured conditioning a unitary operation on the following one \cite{Slicing,Pierobon,Sych}.
\par
In this paper we address single-shot binary discrimination in realistic conditions. In particular, we propose a hybrid scheme based on the combination of the homodyne-like and direct detection and prove it to be robust against detector inefficiencies and phase noise affecting the input signals.
More in detail, we exploit a homodyne setup, that we call \textit{homodyne-like}, where the usual p-i-n photodiodes are replaced with photon-number-resolving detectors (PNR) having a finite photon number resolution \cite{DiMarioBecerra} and a low local oscillator is considered \cite{Allevi,Bina}. In our theoretical analysis, we also include the presence of a quantum efficiency $\eta<1$, of a non-zero dark count rate $\nu$ and a visibility reduction $\xi<1$.
\par
The structure of the paper is the following. In Sec.s \ref{sec: QDT} and \ref{sec: HDlike} we recall the basics of binary discrimination theory and describe the features of homodyne-like detection, respectively. Then, in Sec. \ref{sec: HybridRec} we present our proposal of a hybrid receiver employing both homodyne-like and direct detection schemes and considering ideal PNRs. Finally, in Sec \ref{sec: efficiency} we show the robustness of the hybrid receiver against detection inefficiencies (such as finite resolution of the PNRs, quantum efficiency, dark counts and interference visibility).

\section{Binary discrimination of coherent signals}\label{sec: QDT}
The theory of quantum discrimination between non-orthogonal states has been addressed by Helstrom \cite{Helstrom}. In a general framework a sender encodes a classical symbol ``0" or ``1" onto two quantum states $|\psi_0\rangle$ and $|\psi_1\rangle $ with \textit{a priori} probabilities $q_0$ and $q_1$, respectively. The states are sent through a communication channel and a receiver performs a positive-operator-valued measure (POVM) to infer the encoded values ``0" or ``1". If $p(j|k)$ ($j,k \in\{0,1\}$) is the conditional probability of obtaining the outcome $j$ if $k$ was sent, then the receiver discriminates the states with an error probability $P_{\rm err} = q_0 p(1|0) + q_1 p(0|1)$. The task is to find an optimal POVM that minimizes $P_{\rm err}$ and the corresponding receiver is referred to as \textit{optimum}.
\par
Here we address the discrimination of two pure \textit{coherent states} of a single-mode optical field, that is states of the form $|\zeta\rangle= D(\zeta) |0\rangle$, $\zeta\in\mathbb{C}$, where $D(\zeta)=\exp(\zeta a^\dagger - \zeta^* a)$ is the displacement operator, $a$ being the field operator, $[a, a^\dagger]=1$, and $|0\rangle$ is the vacuum state. 
In particular, we consider a \textit{binary phase-shift-keying} (BPSK) scheme, where the two states to be discriminated are
\begin{equation}
    |\alpha_0\rangle = |-\alpha \rangle \quad \mbox{and} \quad | \alpha_1\rangle = |\alpha \rangle \, ,
\end{equation}
having the same energy $|\alpha|^2$ but opposite phases (a $\pi$ phase shift). In the following we focus on the case of equal \textit{a priori} probabilities $q_0=q_1=1/2$ and, without loss of generality, we assume $\alpha\in\mathbb{R}_+$. 
\par
Helstrom's theory allows to compute the minimum error probability, the corresponding \textit{Helstrom bound}, that reads:
\begin{eqnarray}
\label{eq: HelstromBound}
    P_{\, \rm{Hel}} &= \frac{1}{2} \biggl[ 1- \sqrt{1- 4q_0 q_1 \, |\langle \alpha_0| \alpha_1 \rangle|^2}\biggr]  \\
    &= \frac{1}{2} \biggl[1- \sqrt{1- e^{-4\alpha^2}}\biggr]\, .    
\end{eqnarray}
The optimal measurement strategy achieving such a minimum is the \textit{``cat state" measurement}, defined by the two-valued POVM $\{\Pi_0, \Id-\Pi_0\}$, $\Pi_0 = |\psi_{\rm cat} \rangle \langle \psi_{\rm cat}|$, where $|\psi_{\rm cat} \rangle = c_0(\alpha) |\alpha_0\rangle + c_1(\alpha) |\alpha_1\rangle $ is an optimized ``cat state" \cite{Helstrom}. However, a concrete realization of such a POVM is not an easy task and therefore there exist many alternative feasible schemes. Here we introduce the \textit{Kennedy receiver}, which will be taken as a benchmark throughout the whole paper.
\par
The Kennedy receiver \cite{KennedyR}, also known as \textit{displacement receiver}, consists in the application of a fixed displacement operation $D(\alpha)$ is applied to each of the two pulses sent, with the effect of mapping 
\begin{equation}
    |-\alpha \rangle \rightarrow |0 \rangle  \quad \mbox{and} \quad |\alpha \rangle \rightarrow |2 \alpha \rangle \, .
\end{equation}
This can be seen as a \textit{``nulling operation''} able to send to the vacuum one of the two input signals. The displacement can be implemented by mixing the incoming signals with a properly chosen local oscillator at a beam splitter with a suitable transmissivity. Then, the discrimination problem is turned into vacuum discrimination, which can be performed by employing a on-off detector, leading to the error probability:
\begin{equation}\label{eq: Kennedy}
    P_{\,K} = \frac{|\langle 0 |2 \alpha \rangle |^2 }{2} = \frac{e^{-4\alpha^2}}{2} \, .
\end{equation}
Although $P_{\, K} > P_{\, \rm{Hel}}$, such receiver is \textit{near-optimum}, since in the regime $\alpha^2 \gg 1$ we have $P_{\, K} \approx 2 P_{\, \rm{Hel}}$.
\par
In the following we will consider a generalisation of the Kennedy receiver, to which we will refer as \textit{displacement-PNR receiver} (D-PNR), where the on-off detector is replaced by a photon-number-resolving  (PNR) detector. As will be discussed throughout the paper, the photon number resolution of the detector will turn out to be useful to improve the decision strategy in the presence of realistic inefficiencies of the receiver.

\section{Homodyne-like measurement}\label{sec: HDlike}

\begin{figure}
\begin{center}
\includegraphics[width=0.5\textwidth]{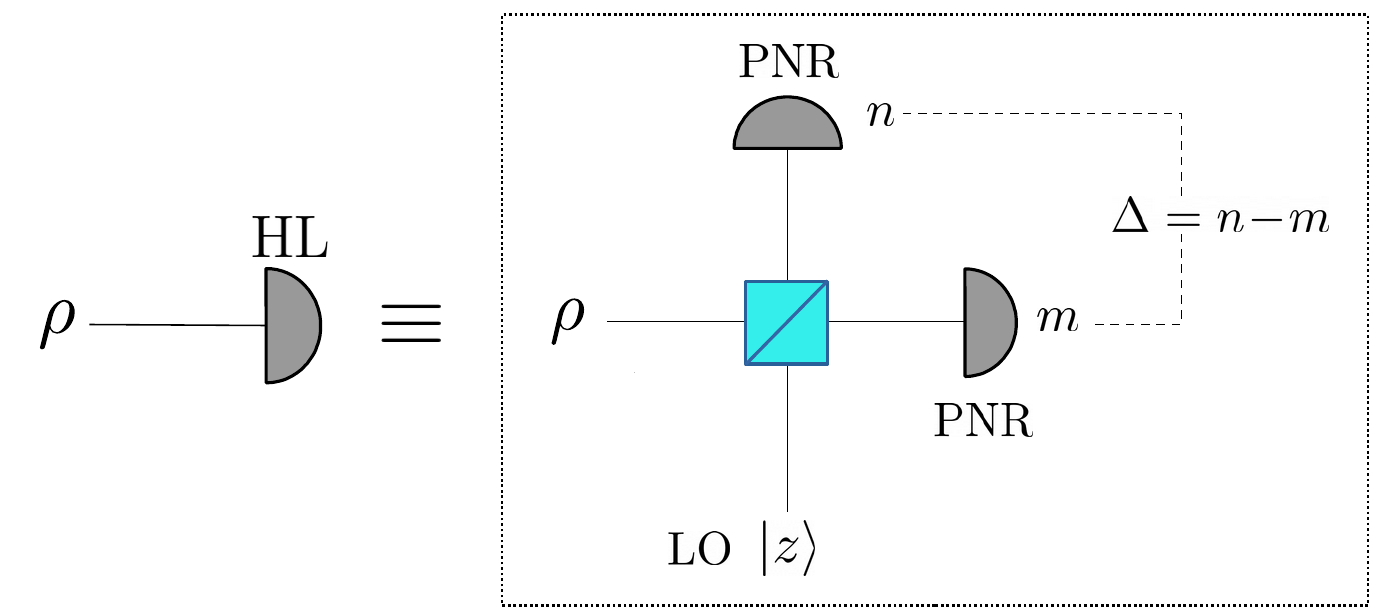} \\[1ex]
\includegraphics[width=0.6\textwidth]{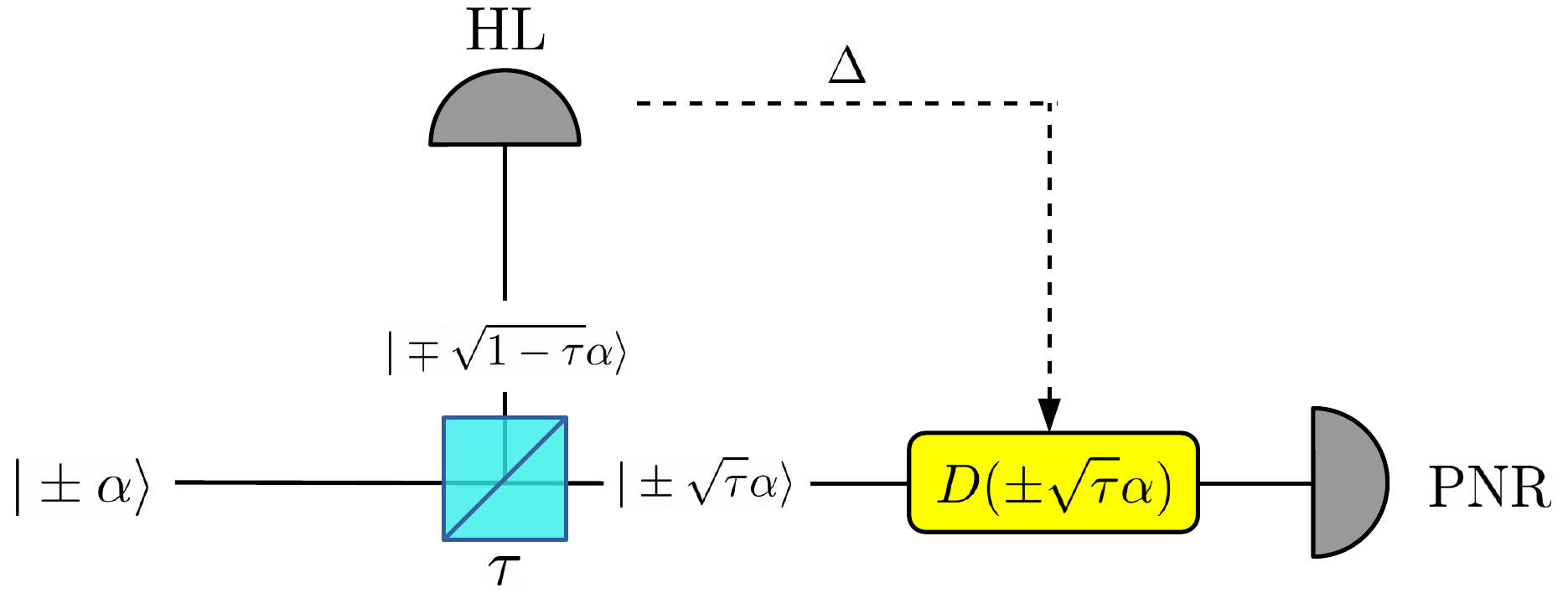}
\end{center}
\caption{(Top) Implementation of homodyne-like detection. The incoming signal is mixed at a balanced beam splitter with a low local oscillator (LO) and then PNR detection is performed on the two branches. (Bottom) Scheme of the hybrid receiver discussed in the paper. The input coherent state is split at a beam splitter of variable transmissivity $\tau$. On the reflected beam we perform homodyne-like detection, whose outcome $\Delta$ conditions a displacement operation on the transmitted signal. After that, we apply on-off measurement on it.}\label{fig:01-Receiver}
\end{figure}

With homodyne-like detection we refer to a homodyne setup which involves photon-number-resolving (PNR) detectors rather than common photodiodes \cite{Allevi}. In this scheme the input state described by the density operator $\rho$ interferes at a balanced beam splitter with a low-intensity local oscillator (LO), prepared in the coherent state $|z\rangle$, $z\in \mathbb{R}_+$.
Then, PNR detection is performed on the beams outgoing the beam splitter, having access to the statistics of the photon numbers $n$ and $m$, respectively. Finally we compute the difference photocurrent $\Delta= n-m$, $\Delta \in \mathbb{Z}$ (see Fig.~\ref{fig:01-Receiver}). 
\par
In the case of our interest,  we consider a coherent input state $\rho = |\zeta \rangle \langle \zeta|$, $\zeta \in \mathbb{C}$. Then, the photocurrent $\Delta$ is the difference of two Poisson random variables and, therefore, follows a \textit{Skellam distribution} \cite{Allevi} :
\begin{equation}\label{eq: Skellam}
    S(\Delta;\zeta) = e^{-\mu_c(\zeta)-\mu_d(\zeta)} \left[\frac{\mu_c(\zeta)}{\mu_d(\zeta)}\right]^{\Delta/2} I_\Delta\Big(2 \sqrt{\mu_c(\zeta) \mu_d(\zeta)}\Big) \ , 
\end{equation}
$\Delta \in \mathbb{Z}$, where 
\begin{equation}\label{eq: mu}
    \mu_{c}(\zeta)= \frac{|\zeta + z|^2}{2},\quad \mbox{and} \quad
    \mu_{d}(\zeta)= \frac{|\zeta - z|^2}{2}
\end{equation}
and $I_\Delta(x)$ is the modified Bessel function of the first kind. 
It is worth noting that, in the regime $z^2 \gg |\zeta|^2$,
\begin{equation}\label{eq: Skellam-HD}
    S(\Delta;\zeta) \rightarrow
\frac{\mathcal{P} (x= \Delta/(\sqrt{2}z); \zeta)}{\sqrt{2}\, z} \, , 
\end{equation}
where:
\begin{equation}\label{eq: HD prob}
    \mathcal{P}(x;\zeta) =\frac{\exp\left[-\left(x - \sqrt{2}\zeta\right)^2\right]}{\sqrt{\pi}}
\end{equation}
is the homodyne probability distribution \cite{OLIrevPLA}.
\par
The scheme described above can be used to implement a \textit{homodyne-like receiver} \cite{Bina}, based on the measured outcome of the photon number difference: if $\Delta <0$ we decide ``0'', if $\Delta >0$ we decide ``1'' and if $\Delta =0$ we perform a random choice. The resulting error probability reads
\begin{equation}\label{eq: HL rec}
    P_{\, HL} = \frac12 \left[ \sum_{\Delta<0} S(\Delta;\alpha) + \sum_{\Delta>0} S(\Delta;-\alpha) \right] + \frac{S_0}{2}\, , 
\end{equation}
with $S_0= S(0;\alpha)= S(0;-\alpha)$.
According to Eq. (\ref{eq: Skellam-HD}), in the limit $z^2 \gg |\zeta|^2$ we regain the traditional \textit{homodyne receiver} whose corresponding error probability reads:
\begin{eqnarray}
    P_{\, H} &= \frac12 \left[\int_0^\inf dx \, \mathcal{P}(x;\alpha_0) +  \int_{-\inf}^0 dx \, \mathcal{P}(x;\alpha_1) \right] \\
    &= \frac{1- \erf\left(\sqrt{2}\alpha\right)}{2}\, ,
\end{eqnarray}
known as standard-quantum-limit (SQL), where $\erf(x)$ is the error function.
\par
In the next paragraph we will see how we can exploit both the direct detection and the homodyne-like receiver to reduce the discrimination error probability. Since the receiver uses at the same time the two detection strategies, we refer to it as \textit{hybrid receiver}.

\section{Near-optimum hybrid receiver}\label{sec: HybridRec}
The scheme of the hybrid receiver proposed in this paper is depicted in Fig. \ref{fig:01-Receiver}. The idea is to exploit a \textit{displacement-PNR} (D-PNR) setup where the nulling displacement is not assigned a priori, but is conditioned on the outcome of a homodyne-like detection performed on a fraction of the input signal.
More in detail, we split the input coherent state $|\alpha_{0/1} \rangle=|\mp \alpha \rangle$ at a beam splitter of variable transmissivity $\tau$ (this can be obtained, for instance, considering the polarization of the input states and by using a polarizing beam splitter), such that:
\begin{equation}\label{eq: BS}
    |\mp \alpha \rangle \otimes |0\rangle
    \rightarrow |\mp \sqrt{\tau} \alpha \rangle \otimes |\pm \sqrt{1-\tau} \alpha \rangle \, .
\end{equation}
Then, we perform homodyne-like detection on the reflected branch
\begin{equation}\label{eq: ReflectedBranch}
    |\alpha^{ (r)}_{0/1}\rangle = |\pm \sqrt{1-\tau} \alpha \rangle \, .
\end{equation}
After that, we apply a feed-forward nulling displacement operation on the transmitted part of the signal conditioned on the outcome $\Delta$ of the homodyne-like measurement:
\numparts \begin{eqnarray}
    &\Delta>0  \rightarrow \mbox{apply } D\left(\sqrt{\tau} \alpha\right) \, , \label{eq: Displacements:a} \\[1ex]
    &\Delta<0  \rightarrow \mbox{apply } D\left(-\sqrt{\tau} \alpha\right) \, , \label{eq: Displacements:b} \\[1ex]
    &\Delta=0  \rightarrow \mbox{apply } D\left(\sqrt{\tau} \alpha\right) \, . \label{eq: Displacements:c}
\end{eqnarray} \endnumparts

Finally, on the resulting displaced state we perform a PNR measurement in terms of on-off detection: the photon number resolution of the detector will turn out to be useful in the presence of dark counts and visibility reduction, as we will see in the following. The intuitive motivation behind the feed-forward rule of Eqs.~(\ref{eq: Displacements:a}), (\ref{eq: Displacements:b}) and (\ref{eq: Displacements:c}) is the following. If we suppose that $|\alpha_0\rangle$ was sent, from the definition of the beam splitter operation of Eq.~(\ref{eq: BS}) it is more likely to obtain $\Delta>0$. As a consequence, we decide to perform a positive displacement sending the transmitted signal into the vacuum such that the PNR detector does not click and we refer to this event as ``off''. Of course there is still a non-zero probability to get $\Delta<0$, and in that case we decide to apply a negative displacement such that the on-off detector is more likely count some photon. This event is called ``on''. Finally, for the case $\Delta=0$, the displacement amplitude is chosen to be positive simply by convention. Analogous considerations may be obtained by considering state $|\alpha_1\rangle$. Given this scenario, the \textit{decision rule} at the end of the final measurement is chosen according to Table~\ref{tab:01-IdealCase}.
\begin{table}
\centering
\begin{tabular}{c |c}
    outcomes & decision \\  \hline
    $\Delta \geq 0$ $\quad$ off \, & ``0" \\ 
    $\Delta < 0$ $\quad$ on \, & ``0" \\ 
    $\Delta < 0$ $\quad$ off \, & ``1" \\ 
    $\Delta \geq 0$ $\quad$ on \, & ``1" \\ \hline
\end{tabular}
\caption{Decision strategy for the hybrid receiver depicted in Fig.~\ref{fig:01-Receiver}.}\label{tab:01-IdealCase}
\end{table}
\par
\begin{figure}
\begin{center}
\includegraphics[width=0.6\textwidth]{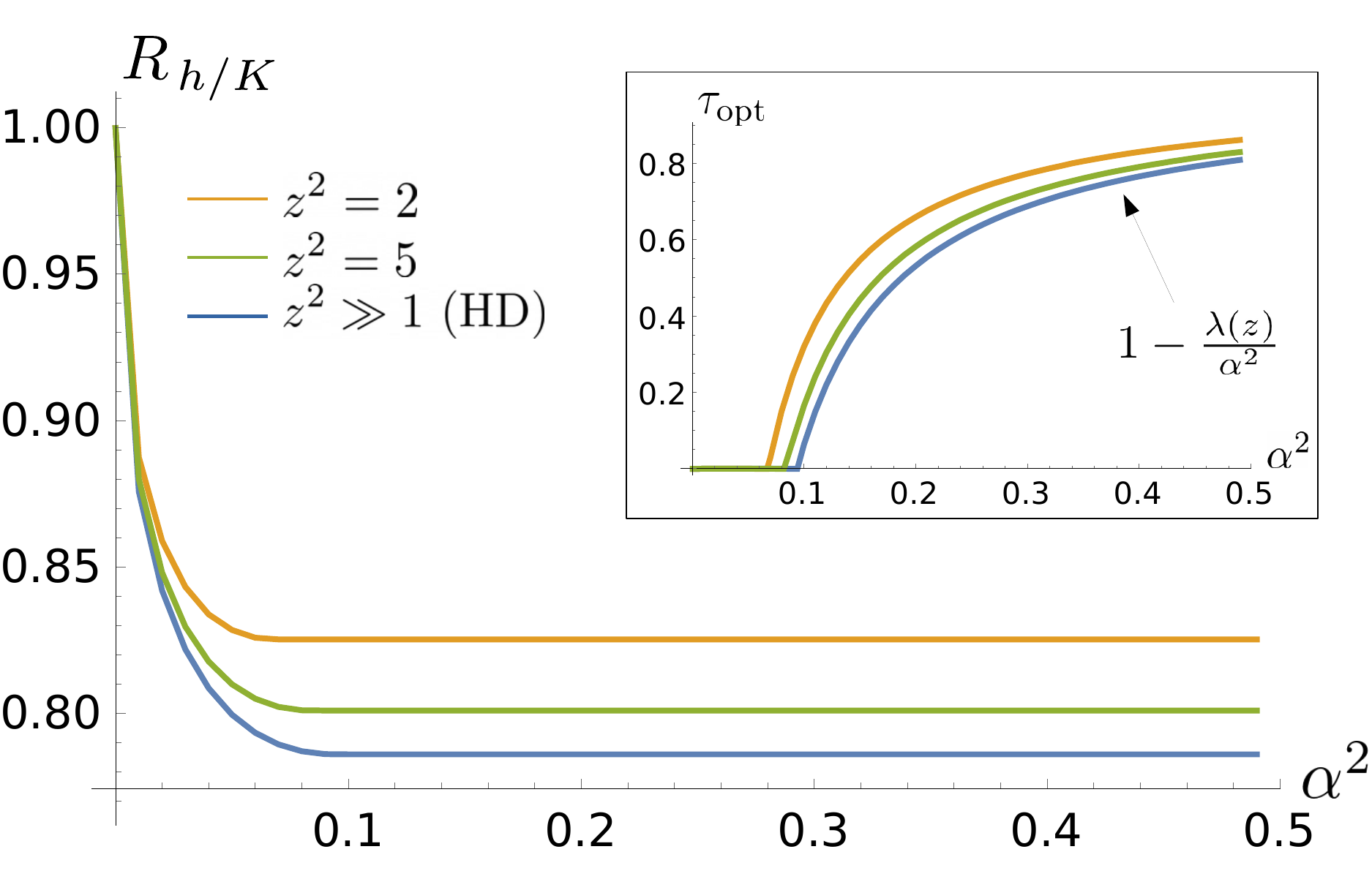}\\[1ex]
\includegraphics[width=0.6\textwidth]{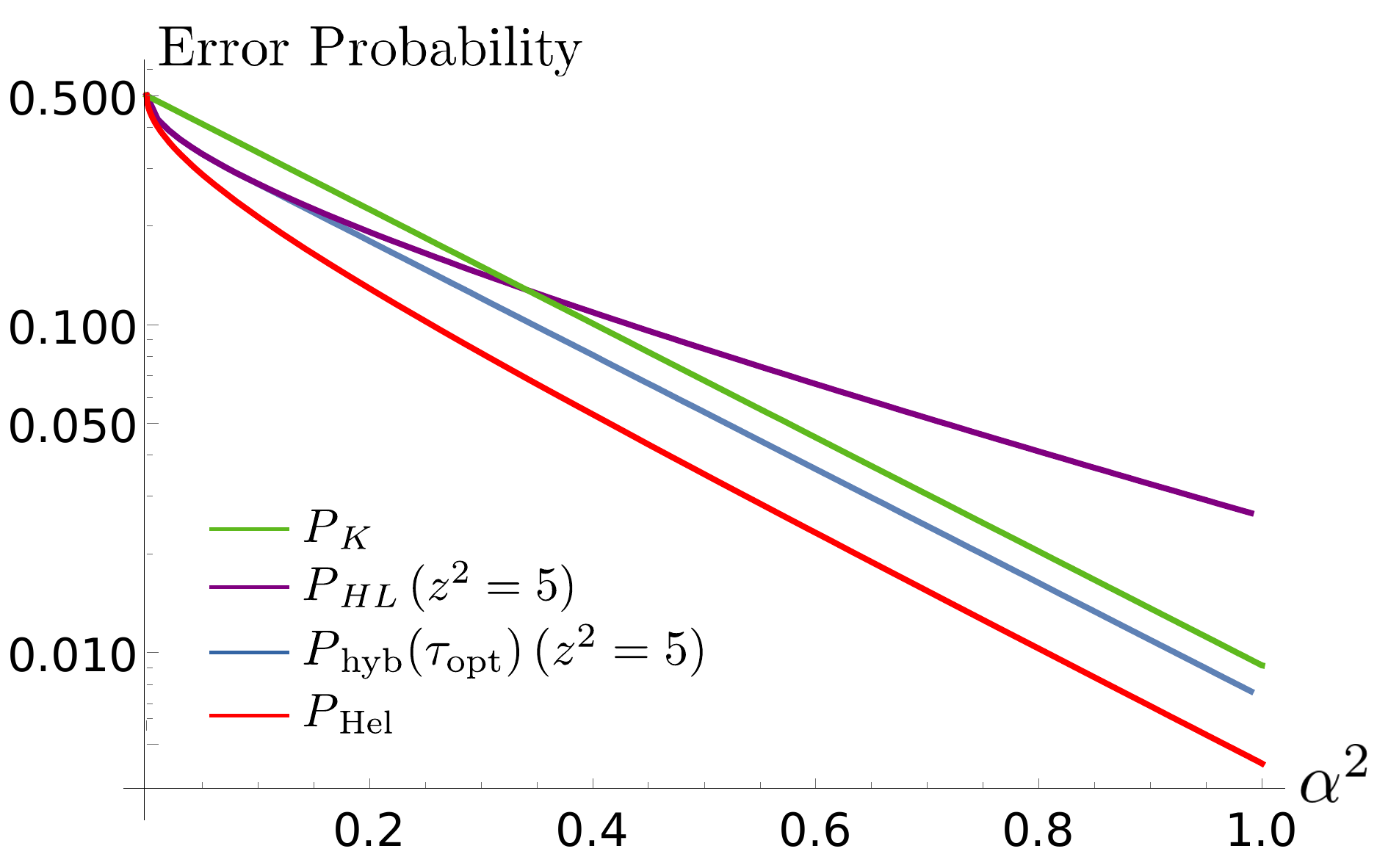}
\end{center}
\caption{(Top) Plot of the ratio $R_{\, h/K}$ as a function of the energy $\alpha^2$ of the encoded pulses for several values of the LO intensity $z^2$. In the inset, plot of the optimized transmissivity $\tau_{\rm opt}$ as a function of $\alpha^2$. For $\alpha^2>N_{\rm th}(z)$ we have $\tau_{\rm opt}= 1-\lambda(z)/\alpha^2$.
(Bottom) Logarithmic plot of the error probabilities as a function of $\alpha^2$ of the proposed hybrid scheme $P_{\, \rm hyb}(\tau_{\rm opt})$ compared to the Kennedy receiver (\ref{eq: Kennedy}), the homodyne-like receiver (\ref{eq: HL rec}) and the Helstrom bound (\ref{eq: HelstromBound}). Here we fix the LO intensity for the homodyne-like receiver and the hybrid receiver to the value $z^2=5$.}\label{fig:02-Hybrid}
\end{figure}

Since:
\begin{equation}
    p(\Delta \geq 0; \mbox{on} | 0)= p(\Delta < 0;\mbox{off} | 1) = 0\, ,
\end{equation}
the error probability for the hybrid receiver is equal to:
\begin{eqnarray}
    P_{\, \rm hyb}(\tau) &= \frac12 \left[ \ p(\Delta < 0; \mbox{off} | 0) +  p(\Delta \geq 0; \mbox{off} | 1)  \right] \nonumber \\[1ex]
    &= \frac12 \left[ \sum_{\Delta < 0} S(\Delta;\alpha^{(r)}_0) e^{-4 \tau \alpha^2}
    + \sum_{\Delta \geq 0} S(\Delta;\alpha^{(r)}_1) e^{-4 \tau \alpha^2} \right] \nonumber \\[1ex]
    &= \frac{e^{-4 \tau \alpha^2}}{2} \left[\sum_{\Delta < 0} S(\Delta;\sqrt{1-\tau}\alpha) 
    +  \sum_{\Delta \geq 0} S(\Delta;-\sqrt{1-\tau}\alpha) \right] \, ,
    \label{eq: pERR}
\end{eqnarray}
where we used the Skellam distribution (\ref{eq: Skellam}). For completeness, we note that performing standard homodyne detection instead of homodyne-like, the error probability of the previous equation becomes
\begin{equation}\label{eq: pERR HD}
    P_{\, \rm hyb}^{\rm (HD)}(\tau) = \frac{e^{-4 \tau \alpha^2}}{2} \ \left\{1- \erf \left[\sqrt{2(1-\tau)} \alpha \right] \right\} \, .
\end{equation}
Equation~(\ref{eq: pERR}) depends on $\tau$, therefore we can optimize it by finding the transmissivity $\tau_{\rm opt}$, that in general is a function of $\alpha^2$, minimizing the value of $P_{\, \rm hyb}(\tau)$ for every $\alpha^2$. Consequently, we obtain the optimized error probability of our receiver $P_{\, \rm hyb}(\tau_{\rm opt})$.
To better enlighten the advantages of the hybrid receiver, it is also relevant to introduce the ratio with the standard Kennedy receiver (\ref{eq: Kennedy}), 
\begin{equation}\label{eq: Ratio}
    R_{\, h/K}= \frac{P_{\,\rm hyb}(\tau_{\rm opt})}{P_{\, K}}\, .
\end{equation}
Plots of $R_{\, h/K}$ and $\tau_{\rm opt}$ are displayed in the top panel of Fig.~\ref{fig:02-Hybrid}, respectively. It emerges that $\tau_{\rm opt}=0$ up to a threshold energy $N_{\rm th}(z)$ which depends on the LO amplitude $z$, while for $\alpha^2 > N_{\rm th}(z)$ it is an increasing function of the energy and reaches asymptotically 1. Therefore, if $\alpha^2 \leq N_{\rm th}(z)$ the optimized strategy is realized with the sole homodyne-like setup, whereas for larger energies the more efficient scheme is obtained by the appropriate interplay between the homodyne-like and the D-PNR parts of our receiver. The choice of the optimal $\tau$ makes the receiver \textit{near-optimum} (see the bottom panel of Fig. \ref{fig:02-Hybrid}) with a ratio $R_{\, h/K}$ saturating to the value $R_{\inf}<1$ for every value of the LO intensity. 

As we noticed, if we increase the intensity of the local oscillator $|z\rangle$, the homodyne-like detection approaches the standard homodyne one. In this case, the ratio in Eq.~(\ref{eq: Ratio}) reads
\begin{equation}\label{eq: Ratio HD}
    R_{\, h/K}^{\rm (HD)}= \frac{P_{\, \rm hyb}^{\rm (HD)}}{P_{K}} = \frac{e^{4 (1-\tau) \alpha^2}}{2} \ \left\{1- \erf \left[\sqrt{2(1-\tau)} \alpha \right] \right\} \, .
\end{equation}
\par
The saturation of $R_{\, h/K}$ for large $\alpha^2$ suggests the following \textit{ansatz} on the expression of the optimized $\tau_{\rm opt}$, namely:
\begin{equation}\label{eq: Ansatz}
    \tau_{\rm opt} = 1- \frac{\lambda(z)}{\alpha^2} \quad \mbox{for} \quad \alpha^2>N_{\rm th}(z) \, ,
\end{equation}
where $\lambda(z)\in {\mathbbm R}_{+}$ and depends on the LO amplitude $z$.
As an example, for the homodyne limit $z^2\rightarrow \inf$, by computing the derivative of Eq.~(\ref{eq: Ratio HD}) with respect to $\tau$ and inserting the expression in Eq.~(\ref{eq: Ansatz}) we get the following relation that must be satisfied by $\lambda \equiv \lambda(z=\inf)$ :
\begin{equation}
    \sqrt{\frac{2}{\pi \lambda}}- 4 e^{2\lambda} \ \Big[1- \erf \bigl(\sqrt{2\lambda} \bigr) \Big] = 0\, ,
\end{equation}
that leads to the numerical solution
$$\lambda \approx 0.094\,.$$
Then, the threshold $N_{\rm th}^{\rm (HD)}\equiv N_{\rm th}(z=\inf)$ can obtained by setting $\tau_{\rm opt}=0$, bringing to $N_{\rm th}^{\rm (HD)}= \lambda$ and the saturation ratio reads:
\begin{equation}
    R_{\inf}^{\rm (HD)} = e^{4 \lambda}  \Bigl[1- \erf \bigl(\sqrt{2\lambda} \bigr) \Bigr] \approx 0.786 \ .
\end{equation}
An identical analysis can be performed for the homodyne-like case, where we may expect $\lambda(z)<\lambda$.

\section{Robustness against detector inefficiencies}\label{sec: efficiency}
To investigate the robustness of our scheme, now we consider a more realistic scheme of the hybrid receiver where we assume to have PNR detectors with a non unit quantum efficiency, dark counts and finite resolution, namely, the detector can resolve up to a given number of photons. Moreover, since the displacement operation is achieved by means of the interference at a beam splitter between the signal and a suitable coherent state, as mentioned above, we also address how a non-unit visibility affects the performance of the receiver.

\subsection{Finite resolution of PNR detectors}\label{sec: PNRM}
\begin{figure}
\begin{center}
\includegraphics[width=0.6\textwidth]{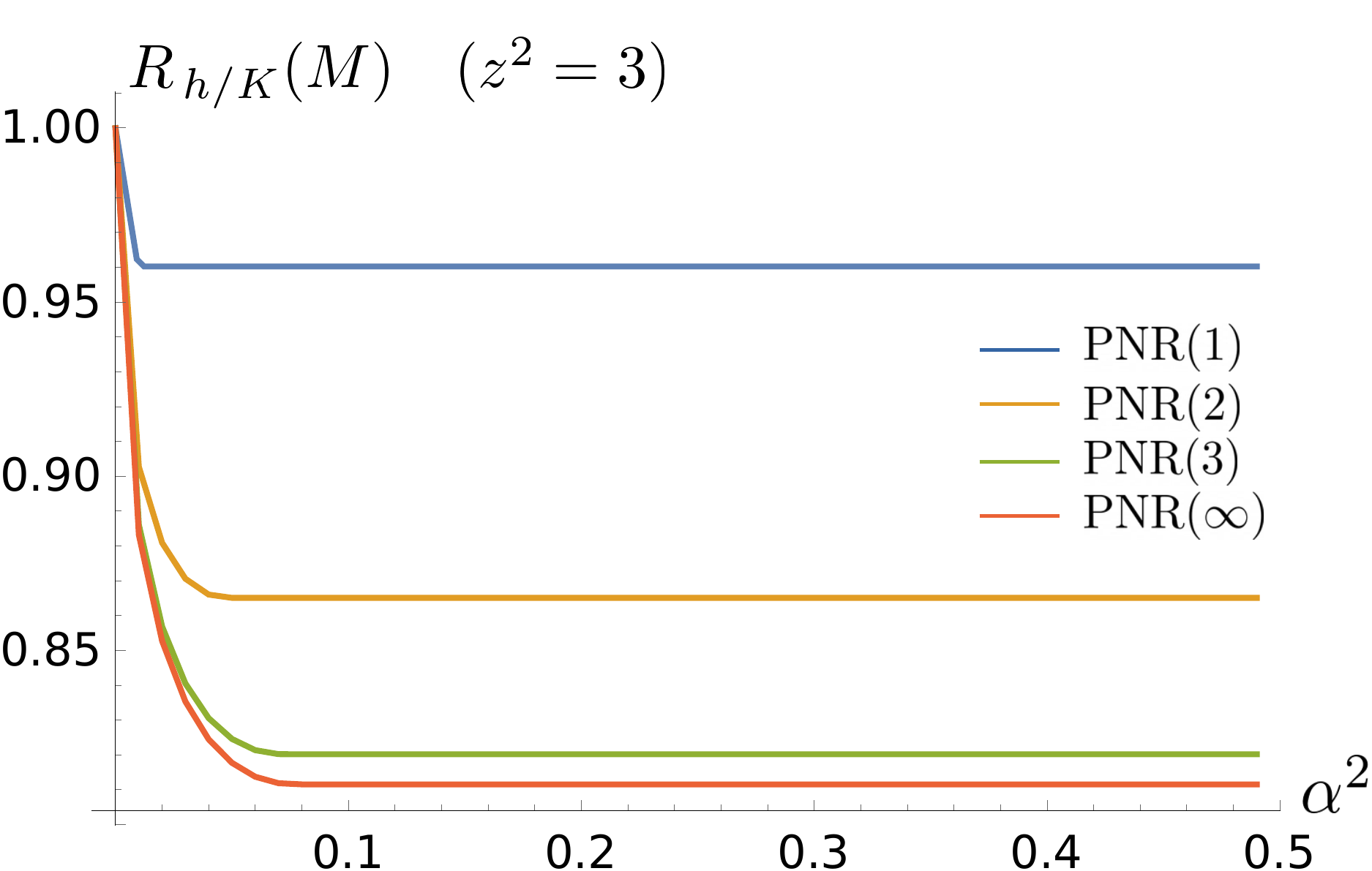}
\end{center}
\caption{Plot of the ratio $R_{h/K}(M)$ as a function of $\alpha^2$ for several values of the PNR resolution $M$. With the notation PNR($\inf$) we refer to the case of ideal PNR. We fix a LO intensity equal to $z^2=3$.}\label{fig:03-PNRM}
\end{figure}

Realistic PNR detectors have a finite photon number resolution, that is they can resolve any number of photons $n$ up to $M$: to highlight this features, we write PNR($M$). For instance, PNR(3) refers to a detectors that has only four possible outcomes $n \in \{0,1,2,\geq\!3\}$, where ``$\geq\!3$'' means 3 or more photons. Clearly, PNR(1) is a on-off photodetector.
\par
PNR($M$) detection may be described through the finite-valued POVM $\{\Pi_n\}_n$, $n=0,...,M$, where:
\begin{eqnarray}
    & \Pi_n = |n\rangle \langle n | \quad \mbox{for} \quad n=0,...,M-1 \, ,\\
    & \Pi_M = \Id - \sum_{n=0}^{M-1} \Pi_n \, .
\end{eqnarray}
As a consequence, if we are performing a PNR($M$) measurement on a generic coherent state $|\zeta\rangle$ ($\zeta \in \mathbb{C}$), the probability of detecting the outcome $n$ reads:
\begin{equation}\label{eq: pn for PNRM}
    p^{(M)}(n;N) = \langle \zeta| \Pi_n |\zeta\rangle =
    \left\{\begin{array}{l l}
    {\displaystyle e^{-N} \ \frac{N^{n}}{n!}}  & n<M \ , \\[2ex]
    {\displaystyle 1- e^{-N} \sum_{j=0}^{M-1} \frac{N^{j}}{j!}} & n = M \ ,
    \end{array}
    \right.
\end{equation}
with a mean photon number $N=|\zeta|^2$.
\par
For the receiver proposed in this paper the exploitation of a PNR($M$) affects the homodyne-like detector. In fact, given Eq.~(\ref{eq: pn for PNRM}) the probability of getting the photon-number difference $\Delta$ after the measurement on the reflected signal \ref{eq: ReflectedBranch} reads
\begin{eqnarray}
    \mathcal{S}&(\Delta;\alpha^{(r)}_{0/1}) =
     \sum_{n=0}^{M} \sum_{m=0}^{M} \delta_{n-m, \Delta} \, p^{(M)}\biggl(n; \mu_c(\alpha^{(r)}_{0/1})\biggr) \
    p^{(M)}\biggl(m;\mu_d(\alpha^{(r)}_{0/1})\biggr) \,  ,\label{eq: pDelta with PNRM}
\end{eqnarray}
$\Delta=-M,...,M$, with the $\mu_{c/d}$ given in Eq.~(\ref{eq: mu}) and $\delta_{k,j}$ is the Kronecker delta. In the limit $M\gg1$, $\mathcal{S}(\Delta;\alpha^{(r)}_{0/1})$ approaches the Skellam distribution (\ref{eq: Skellam}).
\par
The error probability for the hybrid receiver in presence of PNR($M$) is then equal to
\begin{eqnarray}
    P_{\, \rm hyb}^{(M)}(\tau) &= \frac{e^{-4 \tau \alpha^2}}{2} \left[ \sum_{\Delta=-M}^{-1} \mathcal{S}\big(\Delta;\alpha^{(r)}_0\big) 
    + \sum_{\Delta=0}^{M} \mathcal{S}\big(\Delta;\alpha^{(r)}_1\big) \right] \ ,\label{eq: pERR with PNRM}
\end{eqnarray}
which can be optimized to find the transmissivity $\tau_{\rm opt}(M)$, which shows a behaviour qualitatively equivalent to that depicted in Fig. (\ref{fig:02-Hybrid}). The ratio
\begin{equation}\label{eq: R(M)}
    R_{\, h/K}(M)= \frac{P_{\, \rm hyb}^{(M)}\Big(\tau_{\rm opt}(M)\Big)}{P_{\, K}}
\end{equation}
is depicted in Fig.~\ref{fig:03-PNRM}. The effect of the finite resolution is to decrease the saturation ratio $R_{\inf}(M)$, which in any case is still less than $1$, maintaining the advantages of our receiver with respect to the Kennedy.

\subsection{Quantum efficiency $\eta$}\label{sec: QEff}
\begin{figure}
\begin{center}
\includegraphics[width=0.6\textwidth]{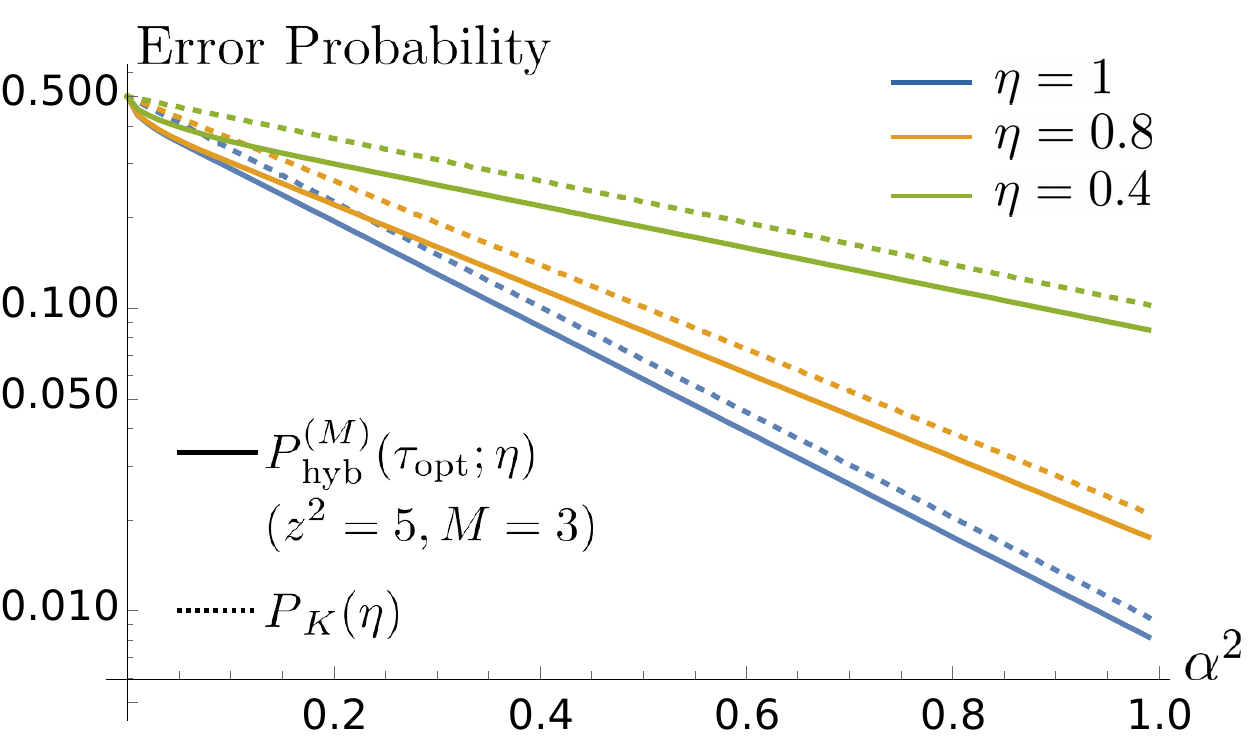}\\[1ex]
\includegraphics[width=0.6\textwidth]{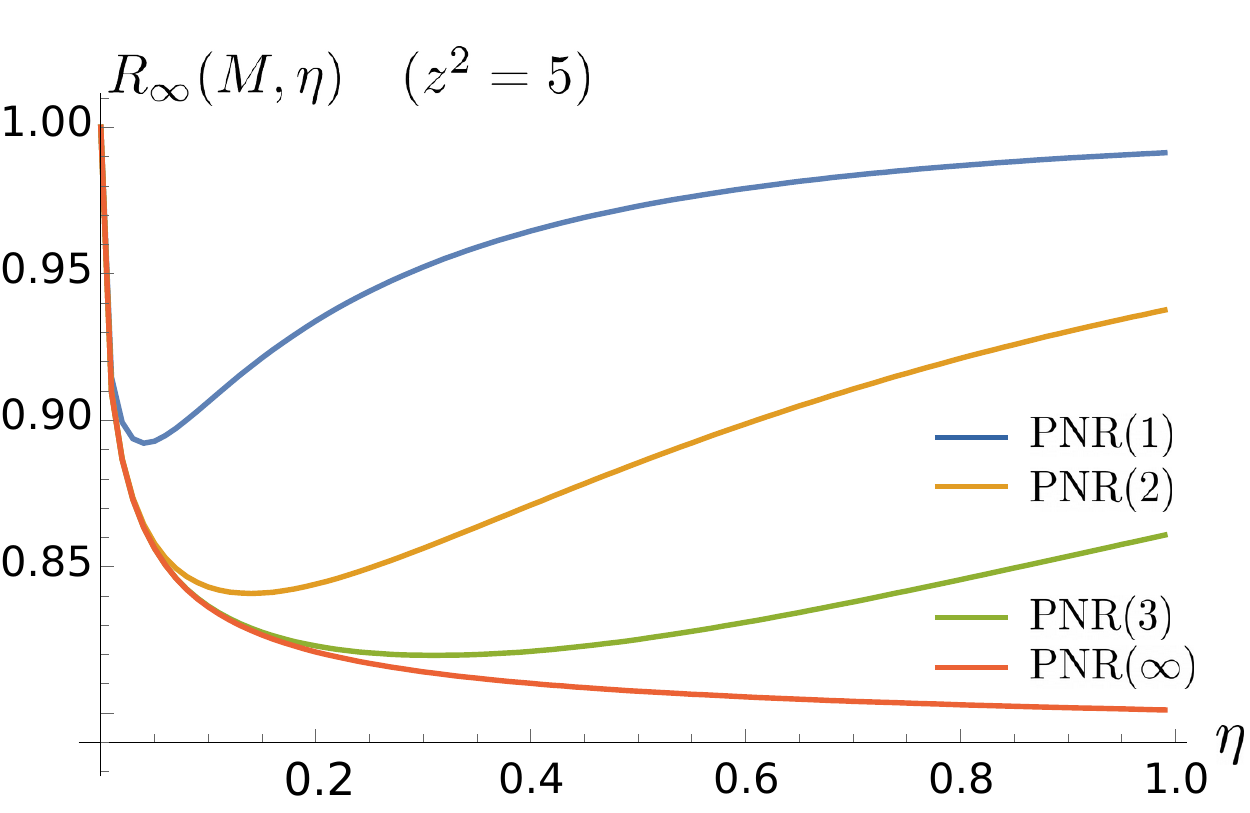}
\end{center}
\caption{(Top) Logarithmic plot of the error probability of the hybrid receiver employing PNR($M$) detectors and the Kennedy receiver as a function of $\alpha^2$ for several values of $\eta$. Here $M=3$. (Bottom) Plot of the saturation ratio $R_{\inf}(M;\eta)$ as a function of the quantum efficiency $\eta$ for several PNR($M$). The LO intensity for both the plots is $z^2=5$.}\label{fig:04-QEFF}
\end{figure}

Concerning the inefficient photodetection, the introduction of a quantum efficiency $\eta$ has the effect of re-scaling all the coherent amplitudes of the measured pulses by a factor $\sqrt{\eta}$, since it corresponds to a photon loss. 
\par
For the \textit{Kennedy receiver} employing inefficient on-off detection, the error probability is changed into:
\begin{equation}\label{eq: KennedyIneff}
    P_{\, K} (\eta) = \frac{e^{-4 \eta \alpha^2}}{2} \, .
\end{equation}
Instead, for the \textit{hybrid receiver} the efficiency affects both the homodyne-like and the PNR measurement schemes. For the homodyne-like detection we have $\mu_c \rightarrow \eta \mu_c$ and $\mu_d \rightarrow \eta \mu_d$, respectively, obtaining 
\begin{eqnarray}
    \mathcal{S}_\eta&(\Delta;\alpha^{(r)}_{0/1}) =
    \sum_{m=0}^{M} \delta_{n-m, \Delta} \, p^{(M)}\biggl(n; \eta \mu_c(\alpha^{(r)}_{0/1})\biggr)\, p^{(M)}\biggl(m;\eta \mu_d(\alpha^{(r)}_{0/1})\biggr) \, .
    \label{eq: pDelta with eff}
\end{eqnarray}
On the other hand, an inefficient on-off detection by the PNR implies the substitution $\exp(-4 \tau \alpha^2) \rightarrow \exp(-4 \eta \tau \alpha^2)$.
By performing these substitutions into Eq.~(\ref{eq: pERR with PNRM}) we get the corresponding error probability:
\begin{eqnarray}
    P_{\, \rm hyb}^{(M)}(\tau; \eta) &=\frac{e^{-4 \eta \tau \alpha^2}}{2} \left[ \sum_{\Delta=-M}^{-1} \mathcal{S}_\eta(\Delta;\alpha^{(r)}_0)
    + \sum_{\Delta=0}^{M} \mathcal{S}_\eta(\Delta;\alpha^{(r)}_1) \right] \, , \label{eq: Phyb with eff}
\end{eqnarray}
and the optimization procedure leads to a different optimized transmissivity $\tau_{\rm opt}(M, \eta)$, which shows the same qualitative behaviour depicted in Fig. \ref{fig:02-Hybrid}. The optimized $P_{\, \rm hyb}^{(M)}(\tau_{\rm opt}(M, \eta); \eta)$ is depicted in the top panel of Fig. \ref{fig:04-QEFF}. For a given value of $\eta$, exploiting the hybrid receiver is still preferable than the Kennedy and the ratio
\begin{equation}\label{eq: R with eff}
    R_{\, h/K}(M,\eta) = \frac{P_{\, \rm hyb}^{(M)}\Big(\tau_{\rm opt}(M,\eta); \eta\Big)}{P_{\, K} (\eta)}
\end{equation}
still saturates to a value $R_{\inf}(M,\eta)$ which depends on $\eta$. The plot of the saturation $R_{\inf}(M,\eta)$ as a function of $\eta$ is depicted in the bottom plot of Fig.~\ref{fig:04-QEFF}. With a resolution $M<\inf$ the function is not monotonic. Indeed, decreasing the value of $\eta$ re-scales the counting rates and reduces the negative consequences induced by the truncation of the Poisson distribution up to $M$. Therefore, if $\eta$ is larger than a given threshold value the ratio $R_{\inf}(M,\eta)$ increases with the efficiency, whereas for smaller $\eta$ the efficiency is too low and $R_{\inf}(M,\eta)$ behaves as a decreasing function.

\subsection{Dark count rate $\nu$}\label{sec: DC}
\begin{figure}
\begin{center}
\includegraphics[width=0.6\columnwidth]{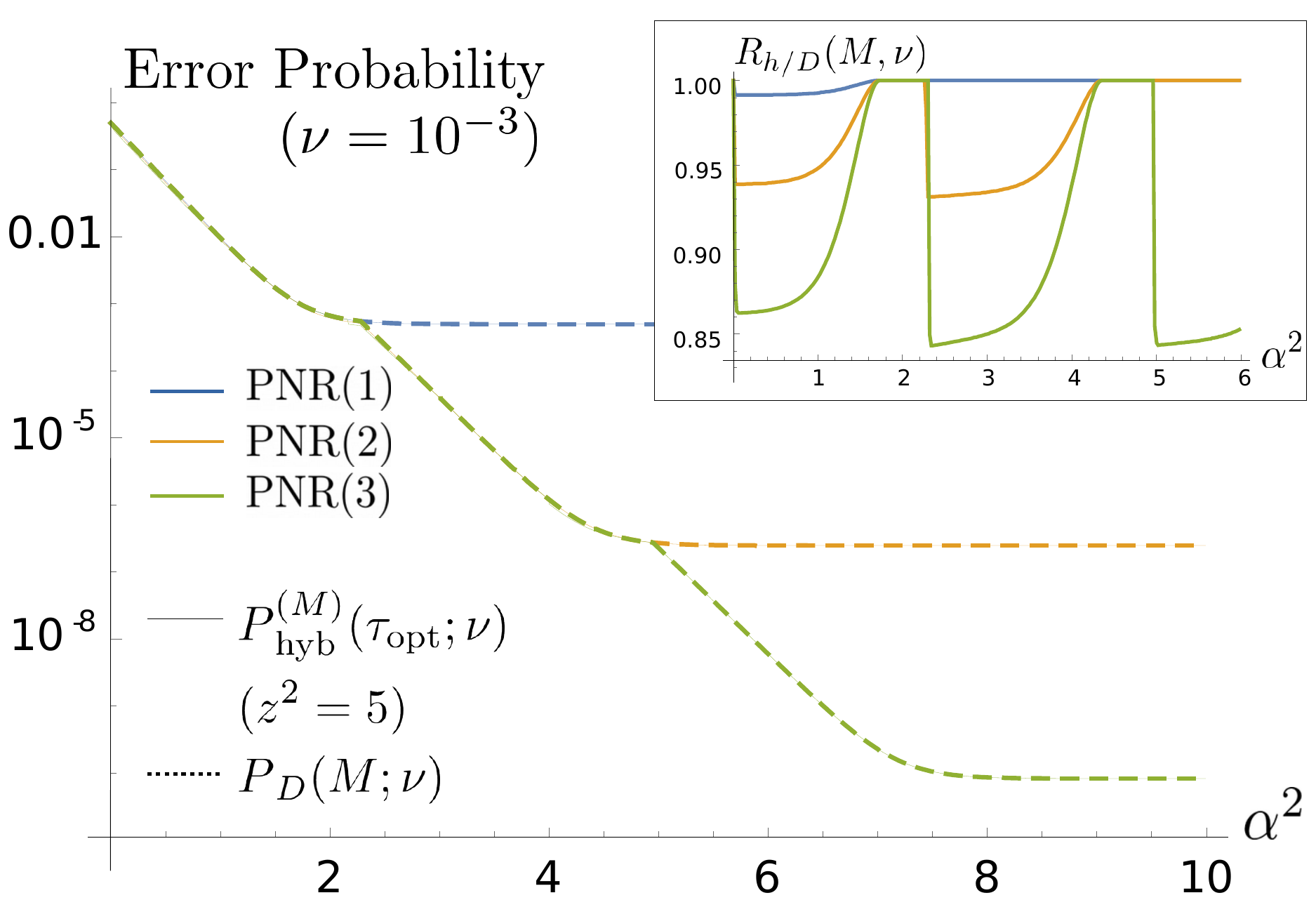}\\[1ex]
\includegraphics[width=0.6\columnwidth]{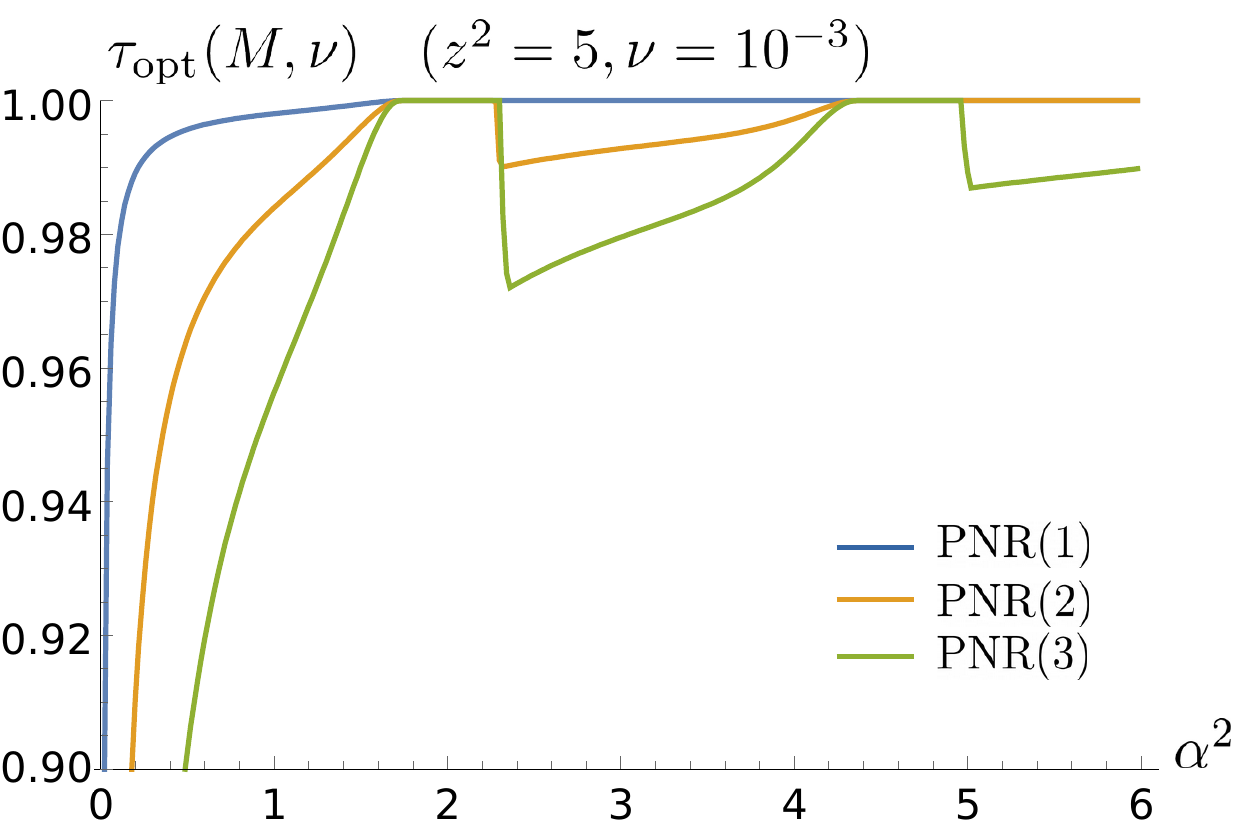}
\end{center}
\caption{(Top) Logarithmic plot of the error probability for the hybrid receiver employing PNR($M$) detectors and the displacement-PNR($M$) receiver as a function of $\alpha^2$ for several values of $M$. In the inset, plot of the ratio $R_{\, h/D}(M,\nu)$ as a function of $\alpha^2$. (Bottom) Plot of the optimized $\tau_{\rm opt}(M,\nu)$ as a function of $\alpha^2$ for several $M$. Here, the dark count rate is set to the value $\nu=10^{-3}$ and the LO for the homodyne-like detector is $z^2=5$.}\label{fig:05- DC}
\end{figure}

Dark counts are random clicks of the PNR due to environmental noise and so not directly correlated to the properties of the coherent measured pulse. Dark counts can be described in terms of Poisson counting \cite{DarkCountsTheory}, occurring at rate $\nu$ which in many realistic conditions takes values $\nu \lesssim 10^{-3}$ \cite{DarkCounts1, DarkCounts2,DarkCounts3,DarkCounts4,DarkCounts5}. Generally speaking, the outcome $n$ of an ideal PNR measurement on a generic coherent state $|\zeta\rangle$ in the presence of dark counts turns out to be the sum of two Poisson variables and, therefore, still follows a Poisson distribution with rate equal to $|\zeta|^2+\nu$ \footnote{The sum of two Poisson independent random variables is still a Poisson random variable. If $x \sim \mathbb{P}(\mu)$ and $y\sim \mathbb{P}(\lambda)$ are two Poisson independent random variables with rates $\mu$ and $\lambda$ respectively, the probability that $x+y$ gets the value $k$ reads $p(x+y=k) = \sum_{l=0}^k p(x=l) p(y=k-l) = e^{-\mu-\lambda} \sum_{l=0}^k \mu^l \lambda^{k-l} /(l! (k-l)!) = e^{-\mu-\lambda} (\mu+\lambda)^k/k! \sim \mathbb{P(\mu+\lambda)}$.}. In the presence of a PNR($M$) we have a probability $p^{(M)}(n;N)$ as in Eq.~(\ref{eq: pn for PNRM}) but with rate $N=|\zeta|^2+\nu$.
\par
The presence of dark counts has a significant effect on the performances of quantum receivers.
In particular, we will now consider as a benchmark the \textit{displacement-PNR($M$) receiver} (D-PNRM) rather than the Kennedy receiver, and exploit the photon number resolution to choose the decision rule for discrimination in a more accurate way. Clearly, the D-PNRM reicever with $M=1$ performs as the Kennedy. Thus, the analysis will proceed in two steps, discussing firstly the cases of D-PNRM and then approaching the hybrid receiver proposed. Without loss of generality in the following we will assume $\eta=1$.

\begin{table}[t]
\centering
\begin{tabular}{c |c}
    outcomes & decision \\  \hline
    $\Delta \geq 0$ $\quad$ $n< n_{\rm th}(\nu)$ \ & ``0" \\ 
    $\Delta < 0$ $\quad$ $n\geq n_{\rm th}(\nu)$ \ & ``0" \\ 
    $\Delta < 0$ $\quad$ $n<n_{\rm th}(\nu)$ \ & ``1" \\ 
    $\Delta \geq 0$ $\quad$ $n\geq n_{\rm th}(\nu)$  \ & ``1" \\ \hline
\end{tabular}
\caption{Decision rule for the hybrid receiver in presence of dark counts.}\label{tab:02-DC}
\end{table}

\paragraph{D-PNRM receiver.} In the presence of dark counts, employing a PNR($M$) detector after the displacement operation rather than a on-off detector brings to some advantages. Indeed, in such a situation the PNR may click even if the vacuum is measured, vanishing the principle behind the nulling technique. As a consequence, the decision rule should be changed according to the \textit{maximum a posteriori probability criterion} (MAP), discussed in \ref{sec: MAP}. If $|\alpha_0 \rangle$ is sent the probability of detecting $n$ photons is $p^{(M)}(n;\nu)$, whereas if $|\alpha_1 \rangle$ is sent the probability is $p^{(M)}(n;4\alpha^2+\nu)$. The error probability for the D-PNRM receiver is then obtained as:
\begin{equation}\label{eq: DPNRM with DC}
    P_{\, D}(M,\nu)= 1- \frac12 \sum_{n=0}^{M} \, \max\Big[p^{(M)}(n;\nu), p^{(M)}(n;4\alpha^2+\nu) \Big] \, .
\end{equation}
The procedure of maximizing the a posteriori probability is equivalent to defining a discrimination threshold $n_{\rm th}(\nu)$ such that all measurement outcomes $n\geq n_{\rm th}(\nu) $ are assigned to state ``1" and all $n< n_{\rm th}(\nu) $ are assigned to state ``0". The threshold number is obtained requiring $p^{(M)}(n_{\rm th};\nu )= p^{(M)}(n_{\rm th};4\alpha^2+\nu)$ and reads
\begin{equation}\label{eq: nTH DC}
    n_{\rm th}(\nu) = \min \left[\left\lceil\frac{4 \alpha^2}{\displaystyle \ln \left(1 + \frac{4\alpha^2}{\nu}\right)}\right\rceil, M \right] \, ,
\end{equation}
where $\ceil{\cdot}$ is the ceiling function. We note that the threshold is a function of $\alpha^2$. For the case of PNR($1$) we have $n_{\rm th}(\nu)=1$, retrieving the on-off discrimination of the standard Kennedy receiver. Plots of the error probabilities for different PNR($M$) are depicted in Fig.~\ref{fig:05- DC} (top panel), where it emerges that dark counts have a drastic effect for large energies, making the error probability saturating. The origin of such effect may be addressed to the finite resolution $M$ of the PNR. Indeed, if $\alpha^2$ is large enough, according to (\ref{eq: nTH DC}) the discrimination threshold will be chosen as $n_{\rm th}(\nu)=M$, thus the sole outcome $M$ will infer state ``1" and all other outcomes smaller than $M$ will infer state ``0". In such a situation the receiver makes the wrong decision only if a $M$ outcome were actually induced by the state $|\alpha_0\rangle$. Then, the error probability for large $\alpha^2$ should be:
\begin{equation}
  P_{\, D}(M,\nu) \approx \frac{p^{(M)}(M;\nu)}{2} = \frac12 \left[1- e^{-\nu} \sum_{j=0}^{M-1} \frac{\nu^{j}}{j!}\right] \, ,
\end{equation}
which is independent on the energy of the pulses $\alpha^2$.

\paragraph{Hybrid receiver.} When considering the hybrid receiver, the presence of dark counts afflicts also homodyne-like detection. Indeed, the probability of obtaining the photocurrent difference $\Delta=-M,...,M$ reads
\begin{eqnarray}
\fl    \mathcal{S}_{\nu}(\Delta;\alpha^{(r)}_{0/1}) = \sum_{n=0}^{M} \sum_{m=0}^{M} \delta_{n-m, \Delta} \, p^{(M)}\bigg(n; \mu_c(\alpha^{(r)}_{0/1})+\nu \bigg)
    p^{(M)}\bigg(m;\mu_d(\alpha^{(r)}_{0/1})+\nu \bigg) \, .
    \label{eq: pDelta with DC}
\end{eqnarray}
Given all the previous considerations, the decision rule for the hybrid receiver in presence of dark counts should be modified into that of Table \ref{tab:02-DC}. 
The error probability then reads:
\begin{eqnarray}
\fl    P^{(M)}_{\, \rm hyb}(\tau; \nu) = q_0 \ \biggl[p(\Delta < 0, n<n_{\rm th}(\nu) | 0) + p(\Delta \geq 0, n\geq n_{\rm th}(\nu) | 0) \biggr] \nonumber\\
+ q_1 \ \biggl[p(\Delta < 0, n\geq n_{\rm th}(\nu) | 1) + p(\Delta \geq 0, n<n_{\rm th}(\nu) | 1) \biggr] \nonumber \\
\fl \textcolor{white}{P^{(M)}_{\, \rm hyb}(\tau; \nu)}  = \frac12 \sum_{n=0}^{n_{\rm th}(\nu)-1} p^{(M)}(n;4\tau \alpha^2+\nu) \left[  \sum_{\Delta=-M}^{-1} \mathcal{S}_{\nu}(\Delta;\alpha^{(r)}_0) + \sum_{\Delta=0}^{M} \mathcal{S}_{\nu}(\Delta;\alpha^{(r)}_1) \right] \nonumber\\
     \hspace{1.cm} + \frac12 \sum_{n=n_{\rm th}(\nu)}^{M} p^{(M)}(n;\nu) \left[  \sum_{\Delta=-M}^{-1} \mathcal{S}_{\nu}(\Delta;\alpha^{(r)}_1) + \sum_{\Delta=0}^{M} \mathcal{S}_{\nu}(\Delta;\alpha^{(r)}_0) \right] \, .
    \label{eq: pERR with DC}
\end{eqnarray}
The optimized error probability $P_{\, \rm hyb}^{(M)}(\tau_{\rm opt}(M,\nu); \nu)$ is depicted in Fig. \ref{fig:05- DC} (top panel), whereas the optimized transmissivity $\tau_{\rm opt}(M,\nu)$ is depicted in the bottom panel. For a better visualization of the advantages brought by the hybrid receiver with respect to the D-PNRM, in the inset of Fig. \ref{fig:05- DC} (top panel) we plot also the ratio 
\begin{equation}
    R_{\, h/D}(M,\nu) = \frac{P_{\, \rm hyb}^{(M)}\Big(\tau_{\rm opt}(M,\nu); \nu\Big)}{P_{\, D}(M,\nu)} \, .
\end{equation}
The behaviour is different from that of Sec. \ref{sec: HybridRec}: at first the value of $\tau_{\rm opt}(M,\nu)$ increases with $\alpha^2$ until to reach exactly the value 1, i.e. performing as a D-PNRM. Accordingly, the ratio $R_{\, h/D}(M,\nu)$ does not saturate but shows a plateau after which increases towards 1. For larger energies, according to the resolution $M$, there appears $M-1$ ``sawteeth", that is other $M-1$ regions in which $\tau_{\rm opt}(M,\nu)$ (and $R_{\, h/D}(M,\nu)$ together with it) decreases to a $<1$ value and increases further to reach again 1. 
Finally, given the results of the previous subsection we note that if a quantum efficiency $\eta<1$ were also present its only effect would be the modification of the plateau value of $R_{\, h/D}(M,\nu)$, preserving the same qualitative behaviour.

\subsection{Visibility $\xi$}\label{sec: Visib}
\begin{figure}
\begin{center}
\includegraphics[width=0.6\textwidth]{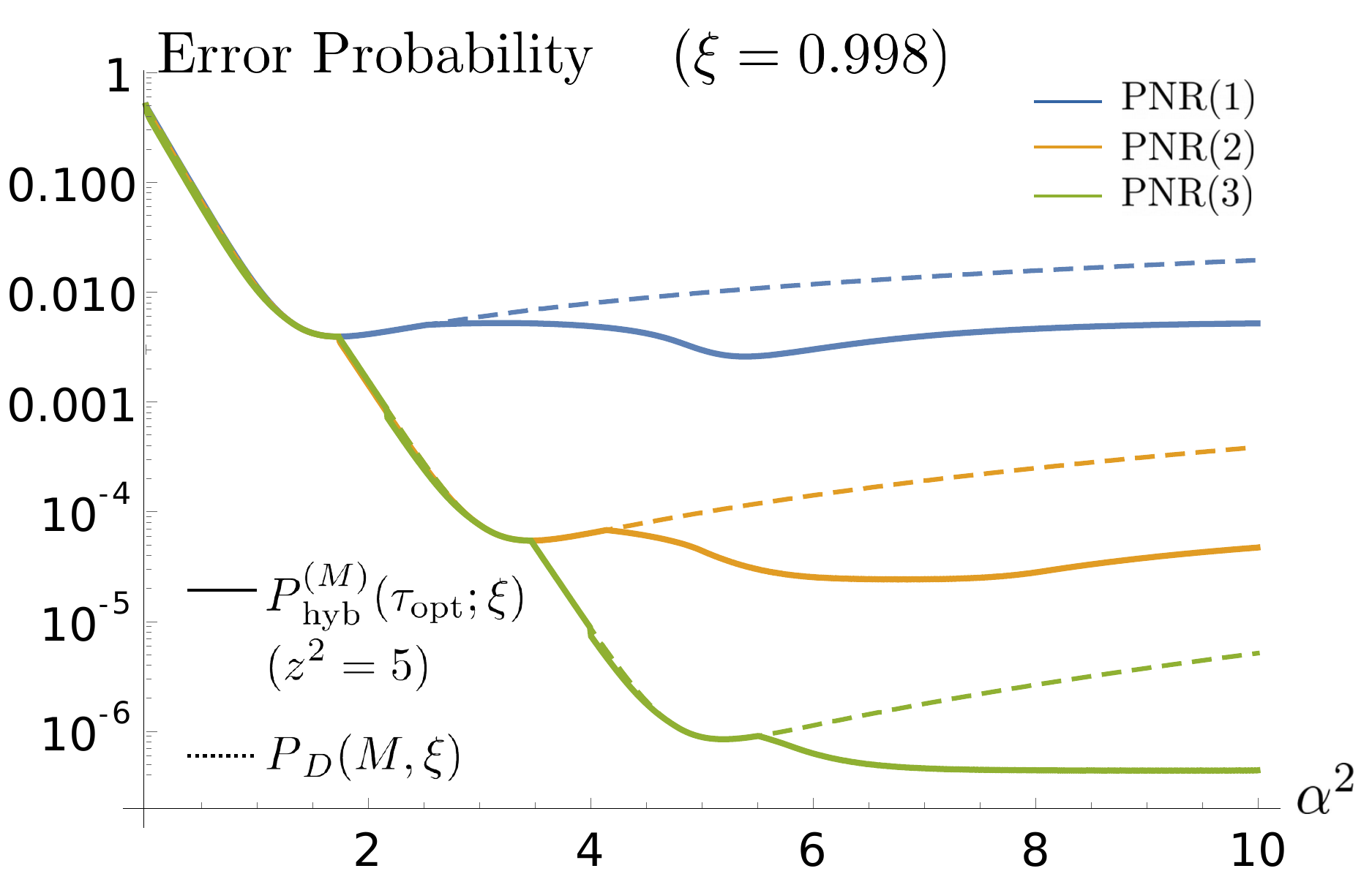}\\[1ex]
\includegraphics[width=0.6\textwidth]{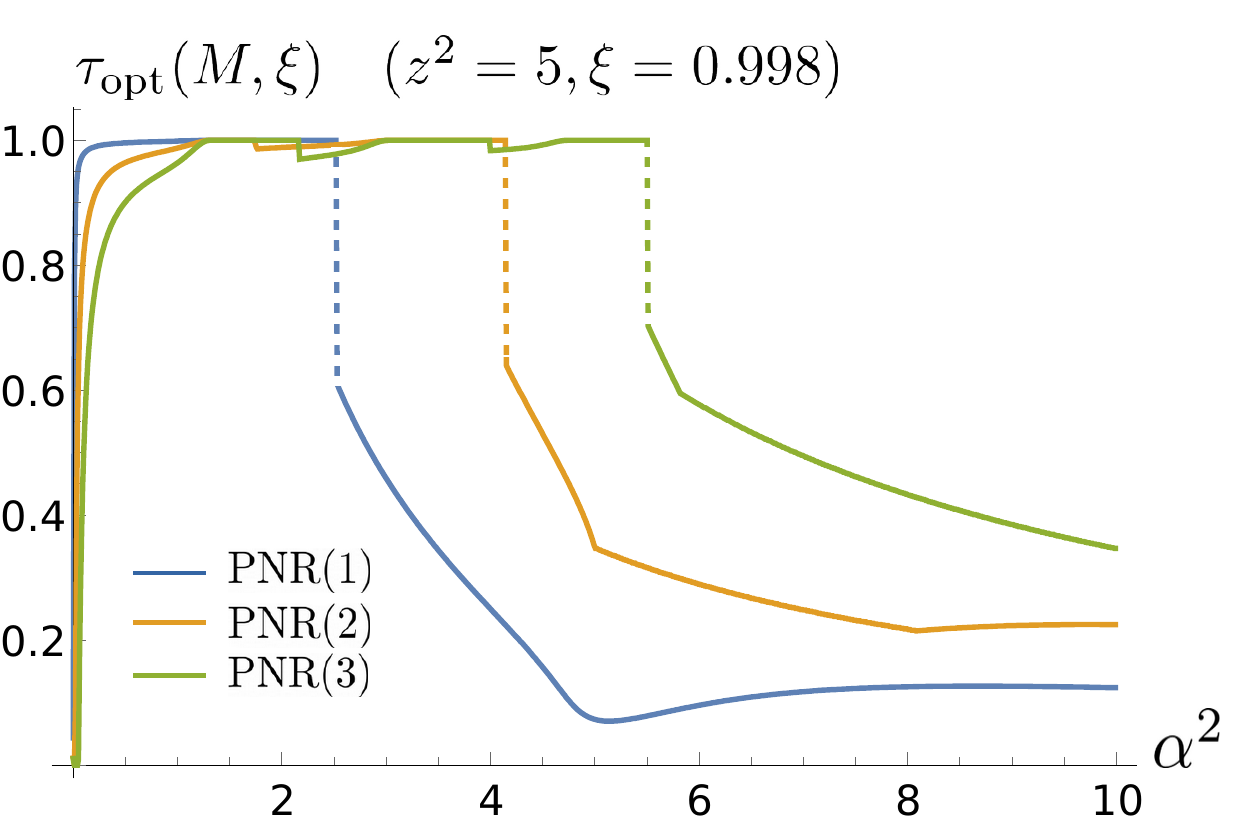}
\end{center}
\caption{(Top) Logarithmic plot of the error probability for the hybrid receiver employing PNR($M$) detectors and the displacement-PNR($M$) receiver as a function of $\alpha^2$ for several values of $M$. (Bottom) Plot of the optimized $\tau_{\rm opt}(M,\xi)$ as a function of $\alpha^2$ for several $M$. Here, the visibility is set to the value $\xi=0.998$ and the LO for the homodyne-like detector is $z^2=5$.}\label{fig:06-Vis}
\end{figure}

Finally, we address the effects of the interference visibility $\xi\leq 1 $ of the displacement operations employed in the realisation of the receiver. This effect is consequence of the mode mismatch at the beam splitter which implements practically a displacement. The value $\xi<1$ quantifies the overlap between the spatial areas of the signal and the auxiliary field mixed at the beam splitter.
As discussed in \cite{DiMarioBecerra,Visibility}, a reduction of the visibility affects crucially the performances of quantum receivers.
\par
Generally speaking, we consider a coherent state $|\zeta\rangle$ which we want to displace by a quantity $\beta$ into the state $|\zeta+\beta\rangle$. For the sake of simplicity, we assume $\zeta,\beta \in \mathbb{R}$. Then we can describe the effect induced by imperfect mode matching by stating that the outcome $n$ of the subsequent PNR measurement follows a Poisson distribution with rate $N= \zeta^2+\beta^2 + 2\xi \zeta \beta \neq (\zeta+\beta)^2$.
As in the previous subsection, we first analyse the cases of D-PNRM receiver and then address the hybrid receiver. As before, we fix $\eta=1$.

\begin{table}[t]
\centering
\begin{tabular}{c |c}
    outcomes & decision \\  \hline
    $\Delta \geq 0$ $\quad$ $n< n_{\rm th}(\xi)$ \ & ``0" \\ 
    $\Delta < 0$ $\quad$ $n\geq n_{\rm th}(\xi)$ \ & ``0" \\ 
    $\Delta < 0$ $\quad$ $n<n_{\rm th}(\xi)$ \ & ``1" \\ 
    $\Delta \geq 0$ $\quad$ $n\geq n_{\rm th}(\xi)$  \ & ``1" \\ \hline
\end{tabular}
\caption{Decision rule for the hybrid receiver in presence of a visibility reduction.}\label{tab:03-Vis}
\end{table}

\paragraph{D-PNRM receiver.} In the presence of a visibility reduction the approach is quite similar to Sec. \ref{sec: DC}. If $|\alpha_0 \rangle$ is sent the probability of detecting outcome $n$ is $p^{(M)}(n;2\alpha^2(1-\xi))$, whereas for $|\alpha_1 \rangle$ the probability is $p^{(M)}(n;2\alpha^2(1+\xi))$. By following the MAP criterion, the error probability then reads
\begin{equation}\label{eq: Kennedy with Vis}
    P_{\, D}(M,\xi)= 1 - \frac12 \sum_{n=0}^{M} \ \max\left[p^{(M)}(n;g_{-}), p^{(M)}(n;g_{+}) \right] \ ,
\end{equation}
where 
\begin{equation}
	g_{\pm} = 2\alpha^2(1\pm\xi) \, ,
\end{equation}
associated to the threshold outcome $n_{\rm th}(\xi)$:
\begin{equation}
    n_{\rm th}(\xi) = \min \left[
    \left\lceil \frac{4 \xi \alpha^2}{\displaystyle \ln \left(1+\xi\right) - \ln \left(1-\xi\right)} \right\rceil, M \right] \ .
\end{equation}
We recall that the case of PNR($1$) is equivalent to the on-off Kennedy receiver. The consequences of a $<1$ visibility on the error probabilities is shown in Fig. \ref{fig:06-Vis}. As for dark counts, the visibility reduction makes the error probability non monotonic, and in particular increasing for large $\alpha^2$. As before, this is a consequence of the finite resolution $M$. In the regime of large $\alpha^2$ the threshold outcome becomes $n_{\rm th}(\xi)=M$, thus the error probability is due to outcomes $M$ induced by the state $|\alpha_0\rangle$ which is not perfectly ``nulled" due to the imperfect displacement operation. Therefore we have:
\begin{eqnarray}
    P_{\, D}(M,\xi) &\approx \frac{p^{(M)}(M;g_{-})}{2} \nonumber \\
    &= \frac12 \left[1- e^{-2\alpha^2(1-\xi)} \sum_{j=0}^{M-1} \frac{\big(2\alpha^2(1-\xi)\big)^{j}}{j!}\right] \, ,
\end{eqnarray}
which is an increasing function of $\alpha^2$.

\paragraph{Hybrid receiver.} For the hybrid receiver, we should also include a visibility reduction in the balanced beam splitter inside the homodyne-like detector. As a consequence, the probability of measuring the photocurrent $\Delta=-M,...,M$ is changed into:
\begin{eqnarray}
 \fl   \mathcal{S}_{\xi}(\Delta;\alpha^{(r)}_{0/1}) &= \sum_{n=0}^{M} \sum_{m=0}^{M} \delta_{n-m, \Delta}\, p^{(M)} \bigg(n; \tilde{\mu}_c(\alpha^{(r)}_{0/1};\xi)\bigg)
p^{(M)}\bigg(m;\tilde{\mu}_d(\alpha^{(r)}_{0/1};\xi)\bigg)  \, ,
    \label{eq: pDelta with Vis}
\end{eqnarray}
where
\numparts \begin{eqnarray}
\label{eq: mu with Vis}
    \tilde{\mu}_c(\alpha^{(r)}_{0/1};\xi)&= \frac{(\alpha^{(r)}_{0/1})^2 + z^2 + 2 \xi \ z \alpha^{(r)}_{0/1}}{2}\ , \\ \tilde{\mu}_d(\alpha^{(r)}_{0/1};\xi)&= \frac{(\alpha^{(r)}_{0/1})^2 + z^2 - 2 \xi \ z \alpha^{(r)}_{0/1}}{2}\, ,
\end{eqnarray} \endnumparts
where $p^{(M)}(n;N)$ is the same of Eq. (\ref{eq: pn for PNRM}).
\par
The decision rule for the hybrid receiver, displayed in Table \ref{tab:03-Vis}, is identical to the case of dark counts.
The error probability then reads:
\begin{eqnarray}
\fl    P^{(M)}_{\, \rm hyb}(\tau; \xi) = q_0 \ \biggl[p(\Delta < 0, n<n_{\rm th}(\xi) | 0) + p(\Delta \geq 0, n\geq n_{\rm th}(\xi) | 0) \biggr]  \  \nonumber\\
     \hspace{1cm}+ q_1\biggl[p(\Delta < 0, n\geq n_{\rm th}(\xi) | 1) + p(\Delta \geq 0, n<n_{\rm th}(\xi) | 1) \biggr] \nonumber \\
\fl \textcolor{white}{P^{(M)}_{\, \rm hyb}(\tau; \xi)} = \frac12 \sum_{n=0}^{n_{\rm th}(\xi)-1} p^{(M)}\big(n;\tau g_{+} \big) \left[  \sum_{\Delta=-M}^{-1} \mathcal{S}_{\xi}(\Delta;\alpha^{(r)}_0) + \sum_{\Delta=0}^{M} \mathcal{S}_{\xi}(\Delta;\alpha^{(r)}_1) \right] \nonumber \\
     \hspace{1cm}   + \frac12 \sum_{n=n_{\rm th}(\xi)}^{M} p^{(M)}\big(n; \tau g_{-}\big) \left[  \sum_{\Delta=-M}^{-1} \mathcal{S}_{\xi}(\Delta;\alpha^{(r)}_1) + \sum_{\Delta=0}^{M} \mathcal{S}_{\xi}(\Delta;\alpha^{(r)}_0) \right] \, .
    \label{eq: pERR with Vis}
\end{eqnarray}
Figure~\ref{fig:06-Vis} shows the optimized $\tau_{\rm opt}(M,\xi)$ and the optimized probability $P_{\, \rm hyb}^{(M)}(\tau_{\rm opt}(M,\xi);\xi)$. If $\alpha^2$ is small we have a behaviour similar to the dark count case, but for large $\alpha^2$ the transmissivity changes discontinuously and the resulting $P_{\, \rm hyb}^{(M)}(\tau_{\rm opt}(M,\xi);\xi)$ keeps always below $P_{\, D}(M,\xi)$. This shows that by choosing appropriately the energy of the signals undergoing the homodyne-like and the D-PNR measurements it is possible to regain part of the information lost by to the finite resolution of the detectors. As a result, the interplay between the two schemes allows to mitigate the negative effects introduced by the visibility reduction.

\section{Conclusions}
In this paper we have advanced the proposal of a new hybrid receiver for binary coherent discrimination, based on the combination of a homodyne-like and Kennedy setups. The incoming signal is split at a beam splitter of variable transmissivity $\tau$, the reflected beam undergoes homodyne-like detection, whose outcome determines a conditioned displacement operation on the transmitted beam, followed by on-off measurement. We have shown that the possibility of adjusting the value of $\tau$ for every value of the energy (for example by exploiting a polarizing beam splitter) makes such receiver near-optimum and capable of beating both the SQL and the Kennedy limit.
\par
Moreover, we have showed that the receiver proves to be robust against the presence of inefficiencies of the experimental setup implementing the receiver, making it a valuable options for realistic experimental implementations of binary receivers. In particular, we have showed that in the presence of a finite resolution $M$ of the PNR detector, an appropriate choice of the transmissivity $\tau$ makes the hybrid receiver beat the performances of the sole displacement-PNR($M$) receiver. Indeed, the possibility of splitting the energy of the coherent seed into two branches allows to regain part of the information lost because of the finite resolution of the detector. 
\par
Further advantages in the regime of small energies may be obtained by following the philosophy of the improved Kennedy receiver \cite{ImprovedKennedy}, that is by optimizing also the amplitude of the displacement operation conditioned on the homodyne-like outcome $\Delta$. By considering an optimized displacement $D(\pm \beta_{\rm opt})$, we expect to maintain the quasi-optimality of the receiver and also to reduce the error probabilities for energies $\alpha^2<1$.

\section*{Acknowledgements}
This work has been partially supported by MAECI, Project No.~PGR06314 ``ENYGMA'' and by University of Milan, Project No.~RV-PSR-SOE-2020-SOLIV ``S--O PhoQuLis''.

\appendix
\section{The maximum a posteriori probability (MAP) criterion}\label{sec:MAP}
We consider a generic displacement-photon counting discrimination scheme to discriminate between the coherent states $|-\alpha\rangle$ and $|\alpha\rangle$ ($\alpha\in\mathbb{R}_+$) generated with equal a priori probabilities $p(\pm\alpha)=1/2$. We apply a fixed displacement of $\beta$ onto the incoming signal, mapping the states into
\begin{equation}
    |\pm \alpha\rangle \rightarrow |\pm\alpha+\beta \rangle \, .
\end{equation}
Then we perform a PNR measurement on the displaced state. The maximum a posteriori probability (MAP) criterion states that, given the outcome $n$, we infer the state with the highest a posteriori probability:
\begin{equation}\label{eq: AP prob}
    p(\pm \alpha | n) = \frac{p(n|\pm \alpha) \ p(\pm\alpha)}{p(n)} \, ,
\end{equation}
where
\begin{equation}
    p(n|\pm \alpha) = e^{-|\pm\alpha+\beta|^2} \frac{|\pm\alpha+\beta|^{2n}}{n!}
\end{equation}
is the probability of getting $n$ photons given $\pm \alpha$ and
$$p(n)= p(\alpha)p(n|\alpha)+p(-\alpha)p(n|-\alpha) = \frac{p(n|\alpha)+p(n|-\alpha)}{2}$$
is the global probability of detecting $n$ photons. For example, we infer $|-\alpha\rangle$ if $p(-\alpha | n)>p( \alpha | n)$, which is equivalent to condition $p(n|-\alpha)>p(n| \alpha)$ since we have $p(\pm\alpha)=1/2$.
\par
The correct decision probability is then equal to
\begin{eqnarray}
    P_c &= p(-\alpha) \sum_{n=0}^\inf p(n|-\alpha) \chi_{-\alpha} + p(\alpha) \sum_{n=0}^\inf p(n|\alpha) \chi_{\alpha} \\
    &= \frac{1}{2} \sum_{n=0}^\inf \max[p(n|-\alpha), p(n|\alpha)] \, .
    \label{eq: pc}
\end{eqnarray}
where
$\chi_{-\alpha}=1$ if $p(n|-\alpha)>p(n| \alpha)$ and 0 otherwise and $\chi_{\alpha}=1$ if $p(n|\alpha)>p(n| -\alpha)$ and 0 otherwise.
The error probability is obtained immediately as $P_{\rm err} = 1- P_c$.
\par
The decision rule $p(n|-\alpha) \lessgtr p(n|\alpha)$ is equivalent to the definition of a threshold outcome $n_{\rm th}$ such that all measurement outcomes $n\geq n_{\rm th} $ are assigned to state $\alpha$ and all $n< n_{\rm th} $ are assigned to state $-\alpha$. The threshold number is obtained by equating $p(n_{\rm th}|-\alpha)= p(n_{\rm th}|\alpha)$ and reads
\begin{equation}
    n_{\rm th} = \ceil[\Bigg]{\frac{|\alpha+\beta|^2-|\alpha-\beta|^2}{\ln \bigl(|\alpha+\beta|^2\bigr) - \ln \bigl(|\alpha-\beta|^2\bigr)}} \ ,
\end{equation}
where $\ceil{x}$ is the ceiling function, returning the smallest integer greater than x.
\par
Finally, we note that for the standard Kennedy receiver the displacement amplitude is $\beta=\alpha$, such that $p(n|-\alpha)= \delta_{n,0}$, therefore the correct probability of Eq.~(\ref{eq: pc}) reduces to $P_c= 1- \exp(-4\alpha^2)/2$.


\section*{References} 

\bibliographystyle{iopart-num}

\providecommand{\newblock}{}
\begin{thebibliography}{}
\expandafter\ifx\csname url\endcsname\relax
  \def\url#1{{\tt #1}}\fi
\expandafter\ifx\csname urlprefix\endcsname\relax\def\urlprefix{URL }\fi
\providecommand{\eprint}[2][]{\url{#2}}

\end{thebibliography}


\begin{thebibliography}{10}
\expandafter\ifx\csname url\endcsname\relax
  \def\url#1{{\tt #1}}\fi
\expandafter\ifx\csname urlprefix\endcsname\relax\def\urlprefix{URL }\fi
\providecommand{\eprint}[2][]{\url{#2}}


\bibitem{Cariolaro}
Cariolaro G 2015 {\em Quantum Communications\/} (Springer Publishing Company,
  Incorporated) 

\bibitem{Helstrom}
Helstrom C~W 1976 {\em Quantum Detection and Estimation Theory\/} Mathematics
  in Science and Engineering 123 (Elsevier, Academic Press)

\bibitem{Bergou}
Bergou J~A 2010 {\em J. Mod. Opt.\/} {\bf 57} 160--180

\bibitem{KennedyR}
Kennedy R~S 1973 {\em Quarterly Progress Report\/} {\bf 108} 219--225

\bibitem{ImprovedKennedy}
Takeoka M and Sasaki M 2008 {\em Phys. Rev. A\/} {\bf 78} 022320

\bibitem{Sasaki&Irota}
Sasaki M and Hirota O 1996 {\em Phys. Rev. A\/} {\bf 54} 2728--2736

\bibitem{Dolinar}
Dolinar S~J 1973 {\em An optimum receiver for the binary coherent state quantum
  channel,} Massachusetts Institute of Technology, Cambridge,
  Technical Report

\bibitem{Slicing}
Takeoka M, Sasaki M and L\"utkenhaus N 2006 {\em Phys. Rev. Lett.\/} {\bf
  97} 040502

\bibitem{Pierobon}
Assalini A, Dalla~Pozza N and Pierobon G 2011 {\em Phys. Rev. A\/} {\bf 84}
  022342 

\bibitem{Sych}
Sych D and Leuchs G 2016 {\em Phys. Rev. Lett.\/} {\bf 117} 200501

\bibitem{DiMarioBecerra}
DiMario M~T and Becerra F~E 2018 {\em Phys. Rev. Lett.\/} {\bf 121} 023603

\bibitem{Allevi}
Allevi A, Bina M, Olivares S and Bondani M 2017 {\em Int. J. Quantum Inf.\/} {\bf 15} 1740016

\bibitem{Bina}
Bina M, Allevi A, Bondani M and Olivares S 2017 {\em Opt. Express\/} {\bf 25}
  10685--10692

\bibitem{OLIrevPLA}
Olivares S 2021 {\em Phys. Lett. A\/} {\bf 418} 127720 

\bibitem{DarkCountsTheory}
Humer G, Peev M, Schaeff C, Ramelow S, Stipčević M and Ursin R 2015 {\em
  Journal of Lightwave Technology\/} {\bf 33} 3098--3107

\bibitem{DarkCounts1}
Izumi S, Takeoka M, Fujiwara M, Pozza N~D, Assalini A, Ema K and Sasaki M 2012
  {\em Phys. Rev. A\/} {\bf 86} 042328

\bibitem{DarkCounts2}
Izumi S, Neergaard-Nielsen J~S and Andersen U~L 2021 {\em PRX Quantum\/} {\bf
  2} 020305

\bibitem{DarkCounts3}
DiMario M~T, Carrasco E, Jackson R~A and Becerra F~E 2018 {\em J. Opt. Soc. Am.
  B\/} {\bf 35} 568--574

\bibitem{DarkCounts4}
Thekkadath G~S, Sempere-Llagostera S, Bell B~A, Patel R~B, Kim M~S and Walmsley
  I~A 2021 {\em Opt. Lett.\/} {\bf 46} 2565--2568

\bibitem{DarkCounts5}
Sidhu J~S, Izumi S, Neergaard-Nielsen J~S, Lupo C and Andersen U~L 2021 {\em
  PRX Quantum\/} {\bf 2} 010332

\bibitem{Visibility}
Becerra F, Fan J and Migdall A 2015 {\em Nat. Photonics\/} {\bf 9} 48--53

\end{thebibliography}

\providecommand{\newblock}{}

\end{document}